\documentclass{sig-alternate-10}

\usepackage[utf8x]{inputenc}

\usepackage{balance}

\usepackage{tikz}
\usetikzlibrary{arrows,shapes,positioning}
\usepackage{hyperref}
\usepackage{xcolor}
\usepackage{bussproofs}
\usepackage{amsmath}
\usepackage{mathabx}
\usepackage{listings}
\usepackage{inconsolata}
\usepackage{algorithm}
\usepackage[noend]{algpseudocode}
\usepackage{bm} 

\usepackage{enumitem} 

\algrenewcommand\algorithmicprocedure{\textbf{proc:}}
\algrenewcommand\algorithmicrequire{\textbf{Precondition:}}
\algrenewcommand\algorithmicindent{1.0em}%
\hypersetup{
    unicode=true,          
    pdftoolbar=true,        
    pdfmenubar=true,        
    pdffitwindow=false,     
    pdfstartview={FitH},    
    pdfauthor={Sebastian Herbst},     
    colorlinks=true,       
    linkcolor=black,          
    citecolor=black,        
    filecolor=magenta,      
    urlcolor=blue           
}

\newfont{\mycrnotice}{ptmr8t at 7pt}
\newfont{\myconfname}{ptmri8t at 7pt}
\let\crnotice\mycrnotice%
\let\confname\myconfname%
\permission{Permission to make digital or hard copies of part or all of this work for personal or classroom use is granted without fee provided that copies are not made or distributed for profit or commercial advantage and that copies bear this notice and the full citation on the first page. Copyrights for components of this work owned by others than the author(s) must be honored. Abstracting with credit is permitted. To copy otherwise, or republish, to post on servers or to redistribute to lists, requires prior specific permission and/or a fee. Request permissions from Permissions@acm.org.}
\conferenceinfo{DEBS’15,} {June 29 - July 3, 2015, OSLO, Norway. \\
{\mycrnotice{Copyright is held by the owner/author(s). Publication rights licensed to ACM.}}}
\copyrightetc{ACM \the\acmcopyr}
\crdata{978-1-4503-3286-6/15/06\ …\$15.00.\\
http://dx.doi.org/10.1145/2675743.2771830}

\begin{document}

\title{Sequences, yet Functions: The Dual Nature of Data-Stream Processing}


\numberofauthors{4}
\author{
    \alignauthor
        Sebastian Herbst\\
        \affaddr{Friedrich-Alexander-Universität Erlangen-Nürnberg (FAU)}
        \affaddr{Computer Science 6}
        \affaddr{Martensstr. 3}
        \affaddr{91058 Erlangen}
        \email{sebastian.herbst@fau.de}
    \alignauthor
        Johannes Tenschert\\
        \affaddr{Friedrich-Alexander-Universität Erlangen-Nürnberg (FAU)}
        \affaddr{Computer Science 6}
        \affaddr{Martensstr. 3}
        \affaddr{91058 Erlangen}
        \email{johannes.tenschert@fau.de}
    \and
    \alignauthor
        Andreas M. Wahl\\
        \affaddr{Friedrich-Alexander-Universität Erlangen-Nürnberg (FAU)}
        \affaddr{Computer Science 6}
        \affaddr{Martensstr. 3}
        \affaddr{91058 Erlangen}
        \email{andreas.wahl@fau.de}
    \alignauthor
        Klaus Meyer-Wegener
        \affaddr{Friedrich-Alexander-Universität Erlangen-Nürnberg (FAU)}
        \affaddr{Computer Science 6}
        \affaddr{Martensstr. 3}
        \affaddr{91058 Erlangen}
        \email{klaus.meyer-wegener@fau.de}
}

\date{\today}

\newcommand{\eg}{e.\,g.}
\newcommand{\ie}{i.\,e.}

\maketitle

\begin{abstract}
    Data-stream processing has continuously risen in importance as the amount of available data
    has been steadily increasing over the last decade.
    Besides traditional domains such as data-center monitoring and click analytics,
    there is an increasing number of network-enabled production machines that generate continuous streams of data.

	Due to their continuous nature, queries on data-streams can be more complex,
    and distinctly harder to understand then database queries.
    As users have to consider operational details, maintenance and debugging become challenging.

    Current approaches model data-streams as sequences, because this is the way they are physically received.
These models result in an implementation-focused perspective.

    We explore an alternate way of modeling data-streams by focusing on time-slicing semantics.
    This focus results in a model based on functions, which is better suited for reasoning
    about query semantics.

    By adapting the definitions of relevant concepts in stream processing to our model, we illustrate the practical usefulness of our approach.
    Thereby, we 
    link data-streams and query primitives
    to concepts in functional programming and mathematics.
    Most noteworthy, we prove that data-streams are monads, and show how to derive monad definitions for current data-stream models.

    We provide an abstract, yet practical perspective on data-stream related subjects
    based on a sound, consistent query model.
    Our work can serve as solid foundation for future data-stream query-languages.
\end{abstract}

%

\keywords{Data-Stream Models,
          Continuous Queries,
          Query Semantics,
          Query Languages,
          Monads,
          Data-Stream Processing,
          Complex Event Processing} 


\newpage
\section{Introduction}
Following a steady increase of available date, data-stream processing has risen to great importance over the last decade.
Data-stream processing is nowadays essential for a multitude of application domains, such as
data-center monitoring \cite{DBLP:conf/btw/DreissigP17} and click analytics~\cite{Akidau15} as well as novel domains like network-enabled production machines~\cite{Bousdekis17, Cammert06}.

The amalgamation of real-world operational details and query semantics is
exemplified by the often cited definition of data streams by Golab and Özsu \cite{Golab03}:
\begin{center}
``A data stream is a real-time, continuous, ordered (implicitly by arrival time or explicitly by time stamp) sequence of items.''
\end{center}
This definition is as much about data-stream processing as it is about data-streams:
\emph{Real-time} is not a specification of what streams are, or what query results to expect,
but a non-functional requirement directed at the implementation.
It refers to the relationship between the availability of a query's input of its output.
\emph{Continuous} states a stream property --- streams are potentially infinite or unbounded ---
and the requirement of partial evaluation. 
The stream property \emph{ordered} refers to the fact, that evaluation semantics depends on the contained time.
It also specifies how to evaluate stream, if no semantically relevant time is available.
\emph{Sequence} refers to the mode of transport and how streams become available over time, as
all models must impose ordering constraints on the contained time stamps.
Otherwise, a real-world system can never know whether it has received all elements concerning a certain time-stamp.

All these aspects are essential in data-stream processing, but not all of them are required to
understand the meaning of queries.
The inherent complexity of queries stems from the manifold association with time.
We write queries over time-associated data, which becomes available over time and
is evaluated over time.
The importance of time is implied by the application domain.

\paragraph{Query Formulation and Maintenance}
From a user perspective, the difficulty in dealing with data-stream queries is caused by several factors:
\begin{description}[itemsep=-2pt]
    \item[Static Data Distribution]
    	Data-streams cannot be\\reorganized as easily as database.
    	Streams originate from sensors or from the machines, which collect the data.
        It is not possible to enforce structuring and constraints, like ETL-phases in data-warehousing, without already engaging in data-stream processing.
    \item[Latency Requirement]
        DSMS are used in la\-ten\-cy\--sen\-si\-tive domains with work on high data rates.
        To minimize latency, real-time automation is very desirable.
        Actions must be performed as soon as possible, which makes manual supervision not 	workable.
    \item[Windows]
        Even the traditional time- and row-based windows have
        differences that are easy to miss~\cite{Botan10}.
        More powerful window-like constructs have been needed in practice~\cite{Maier12, Akidau15}.
        Their exact definitions can be even harder to grasp.
    \item[Bug Reproduction]
        Faulty queries cannot easily be identified by viewing the results, as
        the throughput is huge and faulty behavior might manifest only under conditions,
        which are not satisfied at the point in time when the query is deployed.
        Replaying data after a faulty query has been fixed, might not be possible at all.
        Therefore, informal reasoning is especially important in this context.
\end{description}

Due to the above factors, users of data-stream management systems (DSMS) are more likely to make mistakes.
These mistakes can also be more detrimental than in database contexts.

\paragraph{Problem Statement}
Data streams are often modeled as sequences~\cite{Maier05, Maier12, Akidau15, Abadi03, Abadi05} as this is the way in which the data is physically received.
Sequential stream models interweave meaning and representation,
which often is too concrete to efficiently reason about queries.

This is not how we treat them in most operations.
Minding the sequential character of streams is neither helpful nor necessary to
understand many data-stream operations, \eg\ joins and time-based windows.
Hence, bag-based (multiset-based) models~\cite{Arasu06, Kramer09}, which ignore the sequential character, are employed in more semantics-focused publications.
However, this only removes the sequential aspect but fails to capture the primary mode of access via an ordering attribute.
Consequently, we challenge the traditional modeling of stream as sequences and bags.

\paragraph{Contribution}
We derive a simple, yet flexible model for data-streams and their processing based on established time-slicing semantics.
We prove that data streams can be modeled by monads, which are ubiquitous and well-researched in pure functional programming (FP).
Hence, our model makes comprehensive work about monads readily available for data-stream-specific research (Sec.~\ref{sec:model}).


We illustrate the benefits of the resulting perspective by adapting the definitions of common data-stream-operations to our model.
We show that our model allows to easily decompose these operations into logical units well-known from mathematics and functional language research (Sec.~\ref{sec:discuss}).

We demonstrate the potential of our model to enable further theoretical and practical applications (Sec.~\ref{sec:future}).




\section{Preliminaries}

Before getting into the details of our approach, we provide some preliminary information needed to understand our reasoning.

\subsection{Snapshot-Reducibility}

For our model to be useful, the definitions must match established semantics for streams.
We base our semantics on the definitions of Maier \cite{Maier05} and Kr\"amer et. Seeger \cite{Kramer09},
who specify the semantics of operations on data-streams by mapping them to relational operations using time-slices.

\begin{figure}[h]
\begin{center}\begin{tikzpicture}[->, >=latex]
    \node (sin)  at (0,2) {$S_0,S_1,\ldots,S_n$};
    \node (rin)  at (4,2) {$R_0,R_1,\ldots,R_n$};
    \node (sout) at (0,0) {$S_{out}$};
    \node (rout) at (4,0) {$R_{out}$};
    \path   (sin)  edge node [above] {snapshot} (rin)
        (sout) edge node [above] {snapshot} (rout)
        (sin) edge node [left] {$op_s$} (sout)
        (rin) edge node [left] {$op_r$} (rout)
        ;
\end{tikzpicture}\end{center}
\vspace{-.5cm}
\caption{Snapshot-Reducibility}
\label{fig:snapshot}
\end{figure}
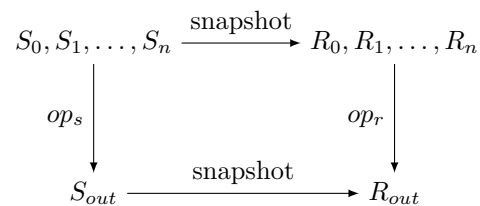

Kr\"amer et. Seeger refer to this concept as snapshot-reducibility. 
It defines the semantics of relational-like operations on streams $op_s$ based on the definitions of the corresponding relational operations $op^r$ (Figure~\ref{fig:snapshot}).
Hence, this does not apply to window operations, which are only defined upon streams and are in general not snapshot-reducible.\footnote{The Now-Window is snapshot-reducible to the identity of relations.}

\subsection{Monads and Comprehensions}

Monad comprehensions have been successfully employed on query constructs in functional programming languages \cite{Wadler90,Giorgidze2011}.
Monads are also used for embedding query constructs in the
.NET-languages under the name of LINQ \cite{Grust10}.
They have proven to be a very useful abstraction in pure, functional programming (FP) overall, and
are thus well investigated (\eg\ \cite{Liang95}).

\subsubsection{Monads}

We do not base our formal representation of monads on category theory whence they originated.
Instead, we base our work on the formulation by Wadler~\cite{Wadler90},
which is geared towards understandability for computer scientists not trained in abstract mathematics.
His work focuses on the structure of monad comprehensions, which are very similar to select statements,
as both were inspired by the mathematical set notation~\cite{Wadler90, Jaeschke85}.
Thus, for our purposes 
a monad is a type-operator $m$ with a triple of polymorphic\footnote{This refers to parametric polymorphism.}
functions\footnote{We renamed $join$ to $flatten$ to prevent confusion with the relational join.}:
\begin{align*}
    map &: \forall x, y:Type~.~(x \to y) \to (m~x \to m~y)\\
    unit &: \forall x:Type~.~ x \to m~x\\
    flatten &: \forall x:Type~.~ m (m ~ x) \to m~x
\end{align*}
\paragraph{Intuition}
As an intuition, consider the set monad ($m = Set $):
$map$ is the element-wise application of a function,
$unit$ creates a singleton set from an element,
and $flatten$ creates a flat set from a set of sets.
Also note, that there is an identity monad ($m~x = x$) where $map$ is function application,
and where $unit$ and $flatten$ are the identity function.

\paragraph{Monad Laws}
As usual for algebraic structures, several laws have to be satisfied.
The laws as stated by Wadler are in lambda-calculus notation, i.e. $f~x$ is function application.
As usual in the functional programming community, we write $\lambda x \to t$ instead of $\lambda x.t$ for lambda abstractions.
We use $\circ$ to denote function composition and $id$ as the identity function $id~x = x$.
\begin{align}
    map~id &= id
    \label{eq:mapid} \\
    map (g \circ f) &= map~g \circ map~f
    \label{eq:mapcomp} \\
    map~f \circ unit &= unit \circ f
    \label{eq:mapunit}\\
    map~f \circ flatten &= flatten \circ (map (map~f))
    \label{eq:mapflatten}\\
    flatten \circ unit &= id
    \label{eq:flattenunit}\\
    flatten \circ map~unit &= id
    \label{eq:flattenmapunit}\\
    flatten \circ flatten &= flatten \circ map~flatten
    \label{eq:flattenflatten}
\end{align}

These laws make extensive use of polymorphism.
For instance law~\eqref{eq:mapid} uses $id$ for
$x$ and for $m~x$.
It could be verbosely written as
$map_{x \to x}~id_x = id_{m~x}$.

\paragraph{Notation}
Like Wadler, we employ overlines to indicate the number of monadic wrappers over a value type.
Therefore, when $x$ is of the element type, which usually is a tuple/record type or XML-tree in data-streams,
$\overline{x}$ is a stream of that element type and $\overline{\overline{x}}$
is a stream of streams of that element type.

Like Wadler, we indicate the monad as a superscript.
$map^{bag}$ is the $map$ function for the bag monad and
$map^{st}$ is the map function for the stream monad.

\subsubsection{Monad Comprehensions}
\label{sec:monadcomp}
As Wadler~\cite{Wadler90} argues, any monad gives rise to a comprehension structure and any comprehension structure gives rise to a monad.
Basic comprehensions allow formulating projections (mappings), singletons and cross products.
In order to provide selections (filtering) a useful definition for the empty comprehension $[~]$ is required.

Giorgidze et al.\cite{Giorgidze2011} provide further desugaring for comprehension,
to realize query constructs from relational databases like grouping and other useful operations.
They require the monad in question to form a monoid in order to use filtering, where $[~]$ is the neutral element.
As we are mainly concerned with set-like structures, we denote the neutral element as $\emptyset$, and the monoid's
associative binary operation $\cup$.
Hence, we write monoid laws as such:
\begin{align*}
    \emptyset \cup a &= a \\
    a \cup \emptyset &= a \\
    (a \cup b) \cup c &= a \cup (b \cup c) \\
\end{align*}
\subsection{Algebraic Datatypes}
\label{sec:algebraic}

We make use of algebraic data types.
The most common type operation is the \emph{product}, which combines two or more types to a new one.
$a \times b$ denotes the product of the types $a$ and $b$ and requires one value of type $a$ and one value
of type $b$ to be constructed.
As usual in set theory, $a \times b$ refers to the type of pairs of $a$ and $b$.

The other, more uncommon type operation in mainstream programming languages is the \emph{sum} type $a + b$ that requires
either a value of type $a$ or a value of type $b$ for construction.
So $a + b$ is the disjoint union of $a$ and $b$ and can be verbally expressed by ``either an $a$ or a $b$''.
In relational database management systems (RDBMS), the constraint NOT NULL actually states that attribute values must be of a certain type $a$.
Otherwise the values are actually of type $a + 1$, where $1$ is any type inhabited by exactly one value, which is named $NULL$ in RDMBS.

\section{Monad-Based Stream Model}
\label{sec:model}

We derive our foundational stream model from snap\-shot-reducibility.
It allows us to give an interpretation of streams of streams.
Then, we expose the monads contained within our model.
Finally, we show how the monad definition can be derived for
arbitrary stream models.

\subsection{Derivation from Snapshots}
\label{sec:dev_snap}

In order to verify snapshot-reducibility, a model for data-streams must come with a definition for snapshots.
More concrete, it is accompanied with a $snapshot$ function that, given a stream and a point in time returns a relation.
However, more often than not, stream models support simultaneous elements, so the word ``relation'' refers
to a bag of tuples instead of a set of tuples.
We go with bags for now, but abstract over this choice in Sec.~\ref{sec:stream_monad}.

We can find many sound stream representations with an accompanying snapshot-function.
To reduce the degrees of freedom, we move the logic of $snapshot$ into
the stream itself to create an abstract model of the concept of a stream.
Like in abstract algebra and in interfaces in programming languages,
a concept is defined by the operations and functions that can be used upon them.

For every custom stream definition $CStream : Type \to Type$, there is a function $snapshot^c$:
\begin{align*}
    snapshot^c : CStream~x \to Time \to Bag~x
\end{align*}
With partial application of any stream $cs: CStream$ to $snapshot^c$ we get a function:
\begin{align*}
    snapshot^c~cs: Time \to Bag~x
\end{align*}
This function contains the content of a stream and the logic of extracting snapshots from said content.
In other words, the simplest model for streams are functions from an arbitrary time domain to
bags of the element type.
The implementation for snapshots in this model $snapshot^s$ collapses to function application:
\begin{align*}
    Stream~x &:= Time \to Bag~x\\
    snapshot^s &= \lambda s \to (\lambda t \to s~t)
\end{align*}

This perspective allows us to focus on the time-sliced semantics.
Note that, if a stream contains no elements, the result is the empty bag.

\subsection{Streams of Streams}

Assessing that streams are monads, incorporates answering the question what streams of streams are,
and finding the correct operational semantics for them.
If streams are modeled as (potentially) infinite sequences,
it is not obvious how to work with infinite sequences of infinite sequences.

Our model provides a more accessible view on what a stream of streams should be, using the above definition
and expanding it:
\begin{align*}
    Stream ( Stream~x) &= Time \to Bag~(Stream ~x) \\
                       &= Time \to Bag~(Time \to Bag~x)
\end{align*}

\paragraph{Interpretation}

A DSMS can be modeled as a function from a stream of queries to a stream of streams of query results.
The input to the system is a stream, since queries may be added and removed over time.
In other words, queries have an associated validity interval.
The results are streams, and these streams themselves inherit the validity interval of their query.
Thus, they are streams themselves.
The content type for all streams involved is an encoding for the network transfer.
For example, we use JSON as a surrogate for an arbitrary network encoding.
A simplistic definition would be:
\begin{align*}
    DSMS := &Stream~JSON \to \\
    & Stream~(Stream~JSON))
\end{align*}
For practical purposes, queries and results are associated with an identifier,
making a more apt definition:
\begin{align*}
    DSMS := & Stream~(\mathbb{N} \times JSON) \\
          &\to Stream~(\mathbb{N} \times Stream~JSON))
\end{align*}

\subsection{Time Domain}

We consider the time type to be an arbitrary, ordered domain of interest.
While some operations, \eg\ row-based windows, require discreteness,
our basic model is agnostic to the choice of time domain, and allows for the level-of-detail needed by users.

However, since we abstract over the sequential nature of streams,
modeling late stream elements requires more advanced time domains as discussed in Sec.~\ref{sec:discuss}.
For the initial modeling, the natural numbers serve well.
\vspace{.2cm}
\subsection{The Stream Monad}
\label{sec:stream_monad}
By viewing streams as functions, the triple of functions to define the stream monad is rather straight-forward,
as function application is the only operation defined on them:
\begin{align*}
    map^{st}~f  &= \lambda \overline{s} \to (\lambda t \to \{f~s| s \in \overline{s}~t \}^{b}) \\
    unit^{st}~s &= \lambda t \to \{ s \}^{b} \\
    flatten^{st}~\overline{\overline{s}} &= \lambda t \to \{ s | s \in \overline{s}~t, \overline{s} \in \overline{\overline{s}}~t\}^{b}
\end{align*}

As mentioned earlier, we do not need to restrict our model to bags.
We can rewrite these functions using the monadic functions of the bag monad following the definition of comprehensions:
\begin{align*}
    map^{st}~f  &= \lambda \overline{s} \to \lambda t \to [f~s | s \gets \overline{s}~t ] \\
               &= \lambda \overline{s} \to \lambda t \to map^{b}~f~(\overline{s}~t) \\
    unit^{st}~s &= \lambda t \to [s] \\
               &= \lambda t \to unit^{b}~s \\
    flatten^{st}~\overline{\overline{s}} &= \lambda t \to [s | s \gets \overline{s}~t, \overline{s} \gets \overline{\overline{s}}~t]\\
               &= \lambda t \to flatten^{b}~(map^{b}~(\lambda \overline{s} \to \overline{s}~t)~(\overline{\overline{s}}~t))
\end{align*}

This form provides a view agnostic to the underlying monad.
Thus, we can define a type function, that
takes a base monad ($b$), a time domain ($t$) and the content type ($x$) as
arguments.
Formally, we define:
\begin{align*}
    BasicStream &: (Type \to Type) \\
    &\to Type \\
    &\to Type \\
    &\to Type \\
    BasicStream~b~t~x &:= t \to b~x
\end{align*}
Hence, the concrete choice of the base monad is irrelevant to this model providing
the freedom to choose set, bags or any other monad as a base depending on the
desired evaluation semantics.
For easier understanding, you may think of the superscript $b$ to mean bag, but indeed it means base.

We have been focusing on the derivation of a model from snapshot-reducibility.
On closer examiniation, we identify the structure of the environment monad~\cite{Liang95} specialized to the time domain.
As our perspective comes from snapshot-reducibility, we call this specialization the $Snapshot$ monad:
\newpage
\vspace{-1cm}
\begin{align*}
    Snapshot~x &= Time \to x\\
    map^{sn}~f &= \lambda \overline{s} \to (\lambda t \to (f~(s~t)))\\
    unit^{sn}~x &= \lambda t \to x\\
    flatten^{sn}~\overline{\overline{s}} &= \lambda t \to \overline{\overline{s}}~t~t
\end{align*}

Effectively, our model is a monad-transformer-stack \cite{Liang95} with the environment transformer on top of the base monad.
Monad-transformer allow stacking of monads to combine them into a monad, which provides all effects of its building blocks.

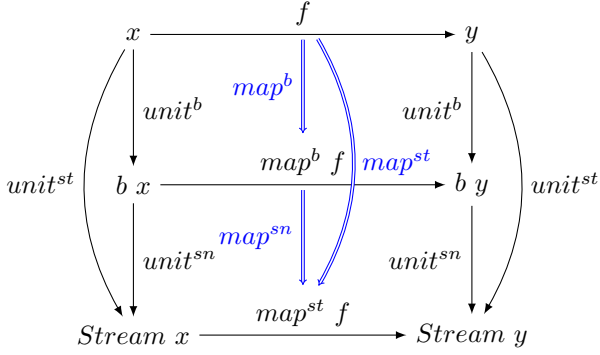
\begin{figure}[h]
\begin{center}\begin{tikzpicture}[>=latex]
    \node (x)  at (0,2) {$x$};
    \node (y)  at (4.5,2) {$y$};
    \node (bx) at (0,0) {$b~x$};
    \node (by) at (4.5,0) {$b~y$};
    \node (sx) at (0,-2) {$Stream~x$};
    \node (sy) at (4.5,-2) {$Stream~y$};
    \path[->] (x) edge node [above] (f) {$f$} (y)
              (bx) edge node [above] (bf) {$map^b~f$} (by)
              (sx) edge node [above] (sf) {$map^{st}~f$} (sy)
              (x) edge node [right] {$unit^b$} (bx)
              (y) edge node [left] {$unit^b$} (by)
              (bx) edge node [right] {$unit^{sn}$} (sx)
              (by) edge node [left] {$unit^{sn}$} (sy)
              (x) edge [bend right] node [left] {$unit^{st}$} (sx)
              (y) edge [bend left] node [right] {$unit^{st}$} (sy)
    ;
    \path[shorten <= 2pt, shorten >= 2pt, -implies, color=blue]
              (f) edge [double] node [left] {$map^b$} (bf)
              (bf) edge [double] node [left] {$map^{sn}$} (sf)
              (f)  edge [double, bend left] node [right] {$map^{st}$} (sf)
              ;

\end{tikzpicture}\end{center}
\vspace{-.5cm}
\caption{Monads in a Stream}
\label{fig:monads}
\end{figure}

We can visualize this stream structure in a commuting diagram as shown in Figure~\ref{fig:monads}.
To keep it readable, we abbreviated $BasicStream~b~t$ with just $Stream$.

\paragraph{Proof}
We provide a formal proof written in Coq\footnote{\url{https://coq.inria.fr/}}, that the monad laws for arbitrary streams are satisfied for any base monad and any time domain (cf.\ App.~\ref{ap:monadlaws}):
\begin{align*}
    \forall b : Type \to Type, Monad~b, t : Type \implies\\ Monad (BasicStream~b~t)
\end{align*}

The correctness is also directly implied by the environment monad-transformer~\cite{Liang95}, but
would require detailed knowledge of monad transformers, whereas the proof we provide only relies on
the preliminaries.

\subsection{Derivation of Monad Definitions for other Models}

As argued in Sec.~\ref{sec:dev_snap}, a stream model is only complete, if it provides a definition for its $snapshot$ function.
Otherwise, the snapshot-reducibility of its operations cannot be verified.

The ability to reconstruct a stream given all snapshots is also highly desirable.
This reconstruction must not be unique, but it should be sound, \ie\ $reconstruct$ must be the right inverse of $snapshot$, but it is not required
to be the left inverse:
\begin{align*}
    snapshot^c &: CustomStream~x \to Time \to Bag~x\\
    reconstruct^c &: (Time \to Bag~x) \to CustomStream~x\\
    snapshot^c &\circ reconstruct^c = id
\end{align*}
This allows for multiple equivalent representations without imposing restriction on implementations, which can
choose elaborate data-structures to achieve high performance.
However, it provides the freedom to use a variety of specialized data-structures in a single system transparently.

These two functions can be used to derive the monadic functions by
replacing function application with $snapshot^c$ in the original definition, and inserting $reconstruct^c$ to
create a custom stream from stream functions:

\begin{align*}
    map^{c}~f  &= \lambda \overline{s} \to reconstruct^c\\
    &(\lambda t \to map^{b}~f~(snapshot^c~\overline{s}~t)) \\
    unit^{c}~s &= reconstruct^c(\lambda t \to unit^{b}~s) \\
    flatten^{c}~\overline{\overline{s}} &= reconstruct^c\\
    &(\lambda t \to flatten^{b}\\
    &(map^{b}~(\lambda \overline{s} \to snapshot^c~\overline{s}~t)~(snapshot^c~\overline{\overline{s}}~t)))
\end{align*}

The above definitions are probably not identical to any original definitions, if given.
However, they are equivalent in terms of snapshot-reducibility, e.\,g.\ $snapshot^c \circ map^c~f = snapshot^c \circ map^{orig}~f$.
If a correct implementation is not available, they can be used as the first, non-optimized implementation.

In general, our view of streams as functions corresponds to an interface specification, whereas models are
an implementation of this interface.
Just as arithmetical functions can be given in various ways, \eg\ closed-form, sets of pairs,
streams can be represented in various ways.
We provide examples for $reconstruct$ for specific models in Sec.~\ref{sec:rel}.

\section{Discussion}
\label{sec:discuss}

In order to illustrate the applicability of our model to data-stream processing, we discuss and investigate relevant concerns in stream processing.
To the best of our knowledge, there is no current and comprehensive taxonomy of issues in data-stream processing available, upon which we can base our discussion.

We opt to discuss aspects ranging from variations of our model and frequently used operations to distributed evaluation.

\subsection{Alternative Base Monads}

In the previous sections, we have mainly used bags as base monad.
However, exploring alternative base monads allows to concisely express restrictions on streams
and the according evaluation semantics.

Every monad (or monad transformer stack) can be used as base monad.
While it may be appealing from a theoretical perspective, constructs like streams of stateful computations and streams of parsers are beyond the scope of data-stream processing.

Hence, we only detail selected base monads relevant to data-stream processing.
We use these in the discussion for operations that cannot be described by snapshot-reducibility.

\subsubsection{Sets}

Choosing sets changes the semantics to automatically remove duplicates.
Relational operations usually do not benefit from duplicates, but duplicates make a difference for most
non-snapshot-reducible operations, \eg\ pattern matching.

However, there is a more subtle difference:
To eliminate duplicates, you must have a definition for equality of elements.
Also for an efficient implementation of sets, you need either an ordering or hashing algorithm.
This is a given, if you think of the element type as tuples of primitive data types.
We do not want to assume the availability of these definitions in general,
as we already encountered streams of streams, where ordering would be expensive.

Instead, we would use sets only if we want to explicitly express this requirement.
However, we did not find corresponding operations requiring this.

\subsubsection{Identity}

The identity monad describes streams, which always contain a singleton value.
Because there is no monoid for $x$ in general, there may be no generic selection.
However, this restriction serves as a guarantee that there is always exactly one value.

Common representatives for this kind are audio streams in waveform.
Mixing two audio signals $\overline{s}_1$, $\overline{s}_2$ would expressed as:
\begin{align*}
    \lambda \overline{s}_1 \to
          \left ( \lambda \overline{s}_2 \to
                \left[ \frac{s_1 + s_2}{2} \middle| s_1 \gets \overline{s}_1, s_2 \gets \overline{s}_2 \right]^{st}
          \right )
\end{align*}

As there is no appropriate definition for $\emptyset$ in the identity monad, selections cannot be performed on audio streams.
Therefore, filtering audio signals is distinctly different operation normal stream filtering.

\subsubsection{Optional Values}

The maybe monad~\cite{Wadler90} $Maybe~x := x + 1$ 
corresponds to the possibility of $NULL$-values in relational database systems.
Instead of bags of values, a single value or none can be valid.
This is what classically is considered an event stream.
At any point in time, there is either an event of a certain type or none.
It provides a neutral element for filtering.
For cross products and joins, there is the choice between a left- or right-biased binary operation.
The maybe monad is a suitable base monad for streams where row-based windows are appropriate (cf. Sec.~\ref{sec:windows}).

\subsubsection{Lists}

As bags can be represented by lists, and the ordering of simultaneous elements may be relevant,
lists can be considered as the base monad.
Although lists are strictly more powerful, this has drawbacks:
Bag union is commutative and thus allows for more algebraic optimization, whereas appending lists is only associative.
However, they are the natural representation of data collected in patterns using the Kleene-Star (cf. Sec.~\ref{sec:pattern}).

\newpage
\subsection{Advanced Time Domains}

Some stream models focus solely on the time stamps provided in the data~\cite{Kramer09, Maier05, Maier12}.
Others take the real-world time into account to model
influences of delays and out-of-order tuples in stream evaluation \cite{Botan10,Akidau15}.
We therefore consider different time domains.
\begin{description}[itemsep=-2pt]
    \item[Event time] The semantically relevant time for applications which is contained in the data. Also referred to as
    application time.
    \item[Arrival time] The time a data becomes available to the DSMS. Also referred to as system time.
    \item[Execution time] The current time at query evaluation in a real stream processor.
\end{description}

To model the influences of arrival time in addition to event time, a product of event and arrival time can be used as the time domain.
As there is no state in our model, the time applied to a result stream expression is propagated to all its input streams.

For instance, to specify that data available more than five seconds late is not to be taken into
consideration for the result, we could add filters accordingly to the input streams, while all other
operations ignore the arrival time:

\begin{align*}
    \lambda \overline{s} \to \lambda (e, a) \to \left\{ \begin{array}{ll}
        \overline{s}~(e, a) & \text{if}\quad a - e \leq 5s \\
        \emptyset & \text{otherwise}
    \end{array}\right.
\end{align*}

If the system allows the update of results such as Dataflow~\cite{Akidau15}, we could also
calculate the correct preliminary results up to a certain arrival time.
An image of the result function of the whole event time domain and the restricted arrival time domain
expresses exactly this.

Modeling execution time is more complex, we, therefore, sketch the approach in Sec.~\ref{sec:future}.

\subsection{Snapshot-Reducible Operations}
\label{sec:stream_comprehensions}

Snapshot-reducible operations are correct by construction.
However, we show some examples for their use in comprehension, as stacked structure provide
a choice at which level comprehensions are used.

\subsubsection{Stream Comprehensions}

As any monad gives rise to a comprehension structure (cf. Section~\ref{sec:monadcomp}),
comprehension can be used on the base monad, the snapshot monad, and the stream monad, which is the combination of the former two (cf. Figure~\ref{fig:monads}).
If used on the stream monad, it specifies stream operations based on \emph{per-element} functions.
If used on the snapshot monad, it specifies stream operations based on \emph{per-bag} or more generally, \emph{per-time-instant}
functions.

As we are not dealing with streams of streams, we omit the overlines, and instead use $s$ for stream variables,
$b$ for base monad variables and $x$ for element variables.

\newpage
\paragraph{Projection}
Projections are usually applied per element in data-stream queries,
and can therefore be expressed using mapping on streams:

\begin{align*}
    proj^{elem} &: (x \to y) \to Stream~x \to Stream~y\\
    proj^{elem} f &= map^{sn}~(\lambda b \to [f~x | x \gets b]^b) \\
                     &= map^{sn}~( map^{b}~f) \\
                     &= \lambda s \to \lambda t \to map^b~f~(s~t) \\
                     &= map^{st}~f\\
                     &= \lambda s \to [f~x | x \gets s]^{st}
\end{align*}
Per time-instance projections allows to change the base monad:
\begin{align*}
    proj^{time} &: (b~x \to b'~y) \to BasicStream~b~t~x \\
                & \to BasicStream~b'~t~y \\
    proj^{time} &= map^{sn} \\
                &= \lambda f \to \lambda s \to [f~b | b \gets s]^{sn}
\end{align*}

\paragraph{Selection}
Because selections cannot be defined on elements, but on the bags or in general the base monad, they are
derived using $map^{sn}$.
However, if selection is defined on the base monad,
we can define selections based on a per-element and on a per-time-instant predicate.

\begin{align*}
    sel^{elem}    &: (x \to \mathbb{B}) \to Stream~x \to Stream~x\\
    sel^{elem}~p  &= map^{sn} (\lambda b \to [x | x \gets b, p~x]^b) \\
    sel^{time}    &: (b~x \to \mathbb{B}) \to BasicStream~b~t~x \\
                  & \to BasicStream~b~t~x\\
    sel^{time}~p~s &= [ x | x \gets s, p~x]^{sn}
\end{align*}

The desugaring for the comprehension filter in both cases requires the definition for $\emptyset^b$ ($[~]^b$ in comprehension syntax).

\subsubsection{Binary Operations}

Deriving n-ary operations from the base monad is a matter of distributing the application of time over the operands; consider the cross product:
\begin{align*}
    (s_1 \times s_2)~t
    &= s_1~t \times s_2~t
\end{align*}

As described by snapshot-reducibility, all operations defined on the base monad can be derived for streams.
Unions, intersections, etc.\ can be derived analogously.

\subsection{Non-snapshot-reducible Operations}

Most data-streams queries contain operations, that are not snapshot-reducible,
most notably windows.
As there is no consent about a complete list, we choose to discuss the most common
operations, and some especially unusual operations.
\newpage
\subsubsection{Traditional Windows}
\label{sec:windows}

As modeled by Maier et al.\cite{Maier05}, window definitions create window instances during evaluation.
Hence, window operations create streams of window instances: They map streams to streams.
As we model streams as functions, window operations are higher-order functions.

\paragraph{Time-Based Windows}

For time-based windows, we can separate the window size and the slide,
and define the highest resolution windows,
\ie\ the windows with the smallest possible slide in the time domain:
\begin{align*}
    &window^\text{time}_{\Delta t} = Stream~x \to Stream~x \\
    &window^\text{time}_{\Delta t} = \lambda \overline{s} \to \lambda t \to \bigcup^b_{t' \in (t - \Delta t; t]} \overline{s}~t'\\
\end{align*}

This can also be written using the image of a function:
\begin{align*}
    &window^\text{time}_{\Delta t}(s,t) = s (t - \Delta t; t]
\end{align*}

Slides specify the desired sampling rate of the resulting stream function\footnote{For a sound definition of a slide, a reference point in time must also be given:
Every full hour after 12:03 is different from every full hour after 12:04.}.
This sampling can be easily specified by filtering all unwanted values.
Given a predicate function $p$ that specifies whether a sample should be available at a certain point in time:
\begin{align*}
    &slide : (\mathbb{T} \to Bool) \to Stream~x \to Stream~x\\
    &slide = \lambda p \to \lambda \overline{s} \to \lambda t \to \left\{\begin{array}{ll}
    \overline{s}~t & \text{if}~p~t\\
    \emptyset^{b} & \text{otherwise}
\end{array}\right.
\end{align*}

Hence, time-based slides are another kind of filter, which uses a predicate on the time opposed to a predicate on the actual data.

\paragraph{Windows for Time-Aware Aggregates}
\label{sec:taa}

We argued \cite{Herbst15} that some aggregates require the window operation to keep
the time information.
To compute the average speed for instance, the time stamps of the measurements are needed.
Otherwise the results are wrong for streams with a variable data-rate or if packets are lost.
Hence, windows for such aggregates must produce instances of type $Time \times x$:
\begin{align*}
    window^\text{taa}_{\Delta t} &: BasicStream~b~t~x \\
                       &\to BasicStream~b~t~(t \times x)
\end{align*}

\paragraph{Row-based Windows}

Row-based windows stem from the perspective of \\streams as sequences,
and can lead to ambiguity when simultaneous elements are allowed, \eg\ the last 5 rows of 7 simultaneous rows.
Akidau et. al\cite{Akidau15} argued that windowing is essentially always
time-based ``over a logical time domain where elements in order have successively increasing logical time stamps.''
This is effectively a restriction on streams to not have simultaneous elements,
which can be expressed by using the identity or the optional/maybe monad as the base monad.
Their typing is pleasantly expressed in our model:
\begin{align*}
    window^\text{row}_n : &BaseStream~Maybe~t~x\\
                          &\to BaseStream~Bag~t~x
\end{align*}

By employing this perspective, another generalization for row-based windows in the presence of simultaneous elements seems
natural.
Instead of counting elements, the operation could count non-neutral content values (\eg\ non-empty bags),
and provide the last $n$ of them:
\begin{align*}
    window^\text{rowgen}_n: &BaseStream~Time~b~x \\
    & \to BaseStream~Time~Bag~(b~x)
\end{align*}

Similar to our modeling of time-slides, row-based slides are a filtering operation, which filters
every result but each $n$th.

\subsubsection{Pattern Matching}
\label{sec:pattern}

Like window operations, pattern matching is not order-agnostic and, hence, not snapshot-reducible.
They consume a stream of events of interest and create a stream of complex events.
As we elaborated \cite{Herbst15}, it is useful to split pattern matching into three logical units:
\begin{itemize}
    \item The actual pattern specifying the \emph{temporal} relationship of different events.
    \item Additional predicates, filtering the results based on the \emph{content}.
    \item The mapping used to create the new complex event type after finding a match.
\end{itemize}
The predicates are selections, the mapping is a projection, and both should be modeled as such.
In real software systems, these operations are usually integrated into the automatons used to identify, filter, and
map matches at once.

Pattern matching means to monitor different event sources continuously to extract temporal sequences of interest.
By creating a sum or disjoint union\footnote{also tagged union} of streams \cite{Herbst15}, a pattern over several streams can be expressed
as a pattern over a sum stream.
As the sum type (cf. Sec.~\ref{sec:algebraic}) appears as the element type, the disjoint union operations
$\sqcup$ can be derived by mapping the sum type's injection functions $in_l$ and $in_r$ over the elements:
\begin{align*}
    \sqcup : &Stream~x \to Stream~y \to Stream (x+y)\\
    s_x \sqcup s_y &= map^{st}~in_l~s_x \cup^{st} map^{st}~in_r~s_y \\
    in_l &: x \to x+y \\
    in_r &: y \to x+y
\end{align*}

The result of the matching operation is a stream of matches.
The type of these matches is implicated by the concrete pattern.
We use regular expression for examples:
\begin{align*}
    &match_{\overline{a}\overline{b}} &:& Stream (a+b+x) \to Stream(a \times b) \\
    &match_{\overline{a}*} &:& Stream~(a+x) \to Stream(List~a) \\
    &match_{\overline{a}|\overline{b}} &:& Stream (a+b+x) \to Stream(a + b) \\
    &match_{\overline{a}?} &:& Stream (a+x) \to Stream (a + 1)
\end{align*}

However, the only limit for patterns is whether matches can be represented in the type system or not.
Adding backtracking, for example, would require changing the implementation, but not the types involved.
In addition, events associated with intervals opposed to time instants require different patterns \cite[pp.~284--286]{Etzion11}.
In general, the input stream's element type is the sum of all event types to be considered, and the output stream's
element type is can be directly constructed from the pattern.

Although regular pattern can be extended to work with simultaneous elements, they
originate from a sequential point of view. To employ this perspective in our model means to
assume the Maybe or Identity monad as base monad.
Then searching for the pattern $aaa$ of $a$s is essentially the same as a row-based window of size three.
The difference is, that row-based windows would create bags with exactly three elements, where
the pattern provides the more specifically typed triple of $a$s, which also retains the ordering information.
Hence, row-based windows are a special case of pattern matching.

\subsubsection{Other Stream Transformations}
\label{sec:inverse}

Abadi et. al \cite{Abadi03} introduced operations which are not snapshot-reducible, but have a different character than
the operations discussed before:
The \emph{resample} operator interpolates a stream, so that the result stream only contains values at the times where reference stream contains values.
It employs windows to specify the scope of interpolation.
If we model the resample operator as a composition of a windowing function, the interpolation calculation and
remaining part $resample = remain \circ interpol \circ window$,
the remaining part has the remarkable property, that is spreads the interpolated data over time.
The same can be said for the \emph{bsort} operator, which effectively works on a row-based window and
fixes a slight disorderliness, which is called slack in their terms.
These operations essentially contain an inverse window operation.

\subsubsection{Aptness}

A generalization for row-based and time-based windows was introduced by Li et.\ Maier et.\ al~\cite{Li05, Maier05}.
They use functions as a means to define windows.
From the data contained in a row\footnote{assuming the current count is also stored in that row},
these functions compute the set of window instances this row belongs to.
Hence, the question of membership in a window instance must be decidable for each row individually.

While this provided more flexibility, Maier et. al.\cite{Maier12} later stated that traditional window constructs are not sufficient.
They introduce the concept of episodes, formalized as T+D-Frames, which target the detection of
burst.
Akidau et. al \cite{Akidau15} introduced session windows which target the same issue, as their instances are terminated
by gaps.

This shows that windows evolved from rather static time- and row-based constructs to more flexible concepts taking more
information about streams into account.
We model them and pattern matching operations as higher-order functions $Stream~x \to Stream~(f~x)$,
where $f$ is an arbitrary type function, which can abstract over them.
But it this approach too generic to accurately reflect the character of windows?
This type schema is broad enough to include the inverse windows identified in Sec.~\ref{sec:inverse}.

Whichever restriction we would want to impose, must be judged by the benefits it provides.
Although the operation incorporating an inverse window aspect seem not popular in research,
we cannot see benefits trying to explicitly exclude them.
The only aspect, we could find to be a reasonable candidate for restrictions, is to require the result stream to not use
values from the future of its input stream.
However, if users wanted to use values from the near future, this would only increase the perceived latency:
Compared to a normal time-based window, the time-base window towards the future, would produce the
same results, associated with time stamp shifted by $(\Delta t - \epsilon)$, where $\epsilon$ is the smallest time difference:
\begin{align*}
    &window^\text{future}_{\Delta t}(s,t) = s[t;t + \Delta t) \\
    &window^\text{time}_{\Delta t}(s,t + \Delta t - \epsilon) = \\
     &\quad s((t + \Delta t - \epsilon) - \Delta t;t + \Delta t - \epsilon] =\\
     &\quad s (t - \epsilon; t + \Delta t - \epsilon] = \\
     &\quad s [t; t + \Delta t) &
\end{align*}
From a practical perspective, this is no issue, whereas requiring values from two weeks ago can well be one.
As this restriction provides no benefits, neither to a stream processor nor to the user,
but would impose restrictions in rare occasions where future values may be requested, we did not pursue this idea.
In conclusion, we strongly doubt that there is a practical, less abstract view on windows.

\subsubsection{Closing Remarks}

Non-snapshot-reducible operations have a great variability.
They can be classified in functional and sequential according to their primary perspective.
For instance, time-based windows are function images and pattern operations search for subsequences.
Session windows~\cite{Akidau15} and episodes~\cite{Maier12} do not fit either category.
They are detect load spikes and gaps.
In order to find them, the sequential and the functional properties must be considered.

\subsection{Distribution}
In real-world application distributed stream processing is an important performance factor.
Hence, we discuss how distribution can be arranged using our stream model.

In contrast to current frameworks, e.\,g.\ Apache Flink~\cite{Carbone2015ApacheFS} and Google Dataflow~\cite{Akidau15}, we do not incorporate distribution keys in our stream model.
Instead, we realize keying can by a function, which projects
the stream content into key-value pairs.
Using a projection function is more flexible then designating an attribute.
In particular, this is an improvement over our previous proposal based on disjoint unions~\cite{Herbst15}.

For any stream of $x$, a function $x \to k \times v$ splitting $x$ into keys $k$ and values $v$ is to be supplied by the user.
We provide a function $partitionWith$ as a sound mechanism that users can employ to explicitly request distribution according to their needs and prior knowledge.
\begin{align*}
    partitionWith : &(x \to k \times v) \to \\
                     & (Stream~v \to Stream~v') \to\\
                     & (k \times v' \to x') \to \\
                     & Stream~x \to ~Stream~x'
\end{align*}

The first argument projects the stream content into pair of keys $k$ and the payload $v$ for keyed evaluation.
The second argument is the stream operation to be evaluated per partition/key and computes values of a possibly different type $v'$.
The third argument is the recombination function, which creates the new elements based on keys $k$ and partitioned stream content $v'$
to create the unified result stream of $x'$.

A distribution function $distribute$ for ordinary key-value pairs can be easily based on $partitionWith$:
\begin{align*}
    distribute : &(Stream (k \times v) \to Stream (k \times v')) \to \\
                 &Stream (k \times v) \to Stream (k \times v')\\
    distribute = &\lambda op \to \\
                 &partititionWith \\
                 & \quad(\lambda (k,v) \to (k,(k,v))\\
                 & \quad op\\
                 & \quad (\lambda (k, (k',z)) \to (k,z))
\end{align*}

As $distribute$ is not a semantics based operation, but rather specifies the request for distributed evaluation, the equation $distribute= id$ is satisfied, unless $op$ changes the key during evaluation which obviously breaks the essential invariance that the keys do not change, \ie\ $k' = k$.
The more generic formulation used in $partitionWith$ is superior to the simple definition of $distribute$,
as it allows the key $k$ to be used in evaluation while enforcing this invariance.



\section{Related Work}
\label{sec:related}
\label{sec:rel}

We discuss our abstraction with respect to established data-stream research.
As we provide a very abstract view, we follow this up by topics from other fields of research,
and thus demonstrate the relationship of data-stream research to them.

\newpage
\subsection{Data Streams}
\label{subsec:datastreams}
Our view of streams as functions is in agreement with Krämer et.\ Seeger's definitions of stream models,
most notably the logical and the physical model.
The logical model is used to define the semantics.
The physical model is used for an efficient implementation of these semantics.

The main feature of the logical model is its uniqueness property: Two stream are equal, iff they are identical.
A logical stream is represented by a set of triples $(e,n,t)$, where
$e$ is the element value,
$n$ is the multiplicity,
and $t$ is the time stamp.
This is analogous to defining a function as a set of pairs.
Hence, the snapshot function for logical stream is:
\begin{align*}
    snapshot_{log}~\overline{s}_{log}~t = \{ (e,n) | (e,n,t') \in \overline{s}_{log} \land t' = t \}
\end{align*}
The reconstruction function is:
\begin{align*}
    reconstruct_{log}~\overline{s} = \{ \overline{s}~t | t \in \mathbb{T} \}
\end{align*}

In the physical model, streams are a bag of pairs of the element value and a validity interval.
This model is used for an efficient implementation of his stream processor.
Equal streams may have different representations, e.g.\ by splitting the validity interval:
\begin{align*}
    \{ (e, [1;5)) \} = \{(e, [1,3)), (e,[3;5))\}
\end{align*}

We do not need to provide a snapshot or a reconstruction function for physical streams, as Krämer
provided conversion functions for physical and logical streams.

Streams in the Dataflow model presented by Akidau et al. \cite{Akidau15} are similar
to physical streams in Krämer's and Seeger's model.
The main differences are in the assumption that every tuple contains a key to be used in parallel
execution, and the usage of compact nested representation for evaluation efficiency.
In addition, the representation switches between time representation as intervals and time instants and
contains interim stages where both are present.
For instance $Droptime stamps$ is an operation to remove time stamps in favor of the contained intervals.
In general, they provide a set of function which have proven its use in practice.
The work is most notable for its modeling of different time domains.
Their approach can handle late tuples by updating the former results.

Maier et al. extended the window concept and provided a formal definition abstracted over the time domain \cite{Maier05}.
Later, they motivated a need for a even broader concept for stream excerpts \cite{Maier12}.
They introduced these \emph{frames} to formally define ``episodes'' of interest,
and showed the insufficiency of traditional window definitions.
Their data model defines streams as a potentially unbounded sequence of tuples.
The event time is referred to it as \emph{progressing attribute} and it is not required
to be monotonically increasing with arrival time.
In fact, the physical ordering or arrival time is viewed solely as a concern for the implementation
and not referred to in the definitions.
As such, their streams may as well by viewed as a set/bag as in Krämer's et Seeger's logical streams,
or more abstractly as functions from time to bags.

In our earlier work \cite{Herbst15}, we focused on a generic set of operations for continuous queries.
We motivated the necessity for pattern matching and more powerful window constructs, and showed
that equality may be proven.
However, we could not provide a formal proof, as our model relied on conventions, and a sequential
perspective is less convenient for proofs than a functional one.
This is one of the main motivations for this paper.

Stanford's CQL\cite{Arasu06} is one of the first \ldots.
All queries are formulated in the SQL-like languages CQL.
Their queries, thus, are basically operations formulated in the bag monad.
They introduce ``relation-to-stream'' operators that specify how create the stream from the results of
repeatedly called relational queries.
This is very similar to mapping bag operations (with $map^{sn}$) to the domain of streams.
However, a result of this bag-focused approach is that they do not provide access to the application time.
As we argued, the time information is required in many real-world scenarios (cf.\ Sec.~\ref{sec:taa}, \cite{Herbst15}).
Our approach of using windows for time-aware-aggregates is an elegant solution,
to provide real-only access, which could also be introduced as an additional window construct in CQL.

\subsection{Streams in Type Theory}

In type theory, streams are considered the categorical dual of lists, \eg\ Harper~\cite[pp.~126--128]{Harper12}.
Hence, in accordance to prior data-stream research the essential modeling of streams is sequential.
A stream is defined by an elimination function, which returns an element and the rest of the stream.
By repeated calling, the stream can be consumed.

The fundamental difference to data-streams is, that streams in type theory are not required to contain time stamps.
However, we can gain the same perspective as Maier et al.~\cite{Maier12} (cf. Sec. \ref{subsec:datastreams}) by adding time stamps to the elimination function, \ie\ :
\begin{align*}
    elimStream : Stream~x \to (x, \mathbb{T}, Stream~x)
\end{align*}

This is dual to a list of pairs of elements and time stamps $List~(x \times \mathbb{T})$, which is the sequential representation of a stream function.

\subsection{Functional Reactive Programming}

Functional reactive programming~\cite{Wan00} (FRP) works on event streams modeled as function fromthe arrival time domain to an arbitrary codomain.
These event functions are either continuous under the name of behaviours, \eg\ the mouse position or the last key pressed,
or discrete under the name of events, \eg\ key released.
Program logic is defined upon the values in the codomain, and
FRP frameworks evaluate the logic, whenever an event function changes their value.
This obviously corresponds to snapshot-reducibility and
most calculations can be implemented as an n-ary cross product followed by a potentially elaborate projection.
In our terms, most event functions are data-streams based on the identity or maybe monad and use the arrival time as the time domain.

%

\section{Future Work}
\label{sec:future}

Our data-stream model has the potential to enable further theoretical and practical applications.
We opt to cover aspects of execution time modeling and implications for future data-stream query-languages.

\subsection{Execution Time}

Execution time concerns can be incorporated into our model.
If execution time is to be considered, the base monad must be replaced with a powerful one
that implements the added semantics:

As real-time constraints are upper-limits, the model has to deal with non-determinism.
Each element has an initial state (before the operation is finished), an ultimate state (after the operation must be finished) and an arbitrary number of intermediate states.

The intermediate states emerge only from the combination of previous operations, as the inputs' arrival time is determined.
This encoding can be facilitated by adding a monadic layer between the base monad and the value type analogous to the non-determinism as shown by Wadler~\cite{Wadler90}.
Monadic encapsulation allows separating the non-determinism from the deterministic operations.

\subsection{Query Languages}

As we demonstrated in Sec.~\ref{sec:stream_comprehensions}, the comprehension syntax is suitable for streams.
However, if users prefer, we can still provide the traditional concrete syntax (\texttt{SELECT-FROM-WHERE}), while benefiting from strong, static typing and type inference.
This means, we can formulate query templates, which work on variables for data-streams and are checked by the stream system without having to know the concrete streams they are applied to.
It also allows for let-polymorphism, which would provide a more powerful \texttt{WITH}.
Nesting can be used without resorting to an XML-representation
making window operations are first-class citizens which can be composed.

In general, a type system alone improves the correctness of queries:
For instance, row-based windows can only be used on streams, which may not have simultaneous elements,
unless our generalization is used.
However, there aspects specific to data-streams, which should get attention:

\paragraph{Schema Reconstruction}

As type inference is solved for monads and comprehension \cite{Jones97}, schema reconstruction is
basically solved.
Interestingly, record types, \ie\ name-indexed tuples, provide the hardest challenge here.
Investigating the best approach for data-stream queries is an open issue.

\paragraph{Automated Distribution}
\label{sec:autodist}
A pure function model allows systems to automatically facilitate distribution, by exploiting equational transformance rules.
In order to do this, it needs to:
\begin{description}
    \item[Identify the attributes] which are invariant during\\parts of the evaluation.
        This can be achieved\\ by formal reasoning on the syntax, if
        the content type is a tuple type.
    \item[Observe aptness of key candidates]
        This can be done by measuring the keys' frequencies during run time, to determine the key which provides the best number of partitions.
    \item[Exploit $id=distribute$].
        This means, replacing subexpressions $se : Stream~(k\times x)\to~Stream~(k\times y)$:
        \begin{align*}
            se~=~&id~se \\
               =~&distribute~se
        \end{align*}
\end{description}
\vspace{-.2cm}
The challenge is to use this information to dynamically invoke the mechanism for distributed query evaluation.

\paragraph{Property Deduction}
\label{sec:properties}

Users may know additional information about the stream sources they used, \eg\ whether they
the stream has a fixed rate, how late elements may be.
The reduction to monads enables the usage of type-inference techniques,
to derived properties of result streams from the properties of the input streams
and additional information about operations.

For performance optimization, and for static analysis (\ie\ type checking),
knowing whether a stream consists of elements associated with time-instants or time spans
is valuable, as the implementation can be chosen accordingly.
For instance, an advertisement-clicked event stream is only associated with a time-instant, \ie\ the time the link was clicked.
We call this an event-stream.
Other streams have data associated with a prolonged activity, like a bus-ride, so the bus-ride-finished event would span from
the start of the ride until the end. We call this a state-stream, as it provides information about a temporary state.
If they are not a direct input to a system, these kinds of streams are generally created by non-snapshot-reducible operations,
as they combine data over a certain time span.
Static data, usually provided as relations, are also used in stream systems, and form a third kind.
We can treat them as streams by using $unit^{sn}$, but this loses valuable information for optimization.

In a sequential model, like Dataflow~\cite{Akidau15}, this character can be reflected by the time stamps,
\ie\ a single time stamp for event streams, an interval for state streams, and the lack of time stamps for static data.
This distinction is essential for the choice of pattern primitives~\cite[pp.~284--286]{Etzion11}.

In a functional model, this must be reflected by a type annotation.
We denote the event tag as $ev$, the state tag as $st$ and the static tag as $et$ for eternity.
Non-snapshot-reducible operations always create state streams that span the time of data collected.

These annotations form a semi-lattice under join op\-er\-ations:
\begin{align*}
    et \bowtie et &= et \qquad &et \bowtie st &= st \qquad &et \bowtie ev &= ev\\
    st \bowtie st &= st &st \bowtie ev &= ev &ev \bowtie ev &= ev\\
\end{align*}

The results are commutative and associative, with the additional laws:
\begin{align*}
    \forall x. \quad ev \bowtie x = ev\\
    \forall x. \quad et \bowtie x = x\\
    st \bowtie st = st\\
\end{align*}

Hence, it should be possible to exploit this during type inference, in order to choose
the most efficient implementation of operations.

\section{Conclusion}

We provided proofs, that data-streams are composed of monads and thus the comprehensive results
dedicated to this algebraic structure can be applied to data-stream processing.
Our notion of data streams as functions from an arbitrary time domain to a base monad provides
a useful abstraction for theoretical studies in this field.
This includes (formal) proofs, like the correctness of algebraic optimizations,
but also support for informal reasoning given a query language based on our ideas.

For real-worlds systems, which certainly employ a more efficient representation of streams,
our definitions can be used to derive correct monadic functions.
As our model abstracts over the choice of the base monad and the time domain,
its users can instantiate it according to their needs.
We expect this approach to lay fertile grounds for query languages with powerful data-stream specific
features while keeping the core simple.

\section{Acknowledgments}
This work has been supported by the Deutsche For\-schungs\-gemeinschaft (DFG) under the grant of FOR 1508 for sub-project no.\ 3.

\bibliographystyle{abbrv}
\bibliography{lib}  

\balancecolumns
\appendix

\section{Proofs}
\label{ap:monadlaws}

On the following pages you can find two coq libraries:
The library \texttt{WadlerMonad} contains the definition of mon-
ads according to Wadler~\cite{Wadler90}. The libary \texttt{StreamMonad}
contains the proof, that streams satisfy the monad laws.

The proofs have been removed, because Arxiv.org is unable to build from
more complex tex sources. The proofs have been generated with a script from Coq sources and
then included in this file.

\end{document}